\documentclass[superscriptaddress,twocolumn,prl]{revtex4}
\usepackage{amssymb}
\usepackage{natbib}
\usepackage{amsmath}
\usepackage{amsfonts}
\usepackage{graphicx}
\usepackage{subfigure}
\usepackage{mathrsfs}
\usepackage{dcolumn} 
\usepackage{bm}
\usepackage{color}
\usepackage{cancel}
\usepackage{mathbbol}
\usepackage{epsfig}
\usepackage{units}
\usepackage{esint}
\usepackage{soul} 
\usepackage{url}
\usepackage{hyperref}

\definecolor{rot}{rgb}{0.75,0.05,0.25}
\definecolor{hellgrau}{gray}{0.5}
\definecolor{blau}{rgb}{0,0,0.7}

\newcommand{\av}[1]{\langle#1\rangle}
\newcommand{\var}[1]{\text{Var}(#1)}

\DeclareMathOperator{\csch}{csch}

\def\Tr{\mbox{Tr}}

\begin{document}

\title[]{Improved bound on entropy production in a quantum annealer}

\author{Michele Campisi}
\affiliation{NEST, Istituto Nanoscienze-CNR and Scuola Normale Superiore, I-56127 Pisa, Italy}
\address{Department of Physics and Astronomy, University of Florence, I-50019, Sesto Fiorentino (FI), Italy}

\author{Lorenzo Buffoni}
\address{Department of Physics and Astronomy, University of Florence, I-50019, Sesto Fiorentino (FI), Italy}

\begin{abstract} 
For a system described by a multivariate probability density function obeying the fluctuation theorem, the average dissipation is lower-bounded by the degree of asymmetry of the marginal distributions (namely the relative entropy between the marginal and its mirror image). We formally prove that such lower bound is tighter than the recently reported bound expressed in terms of the precision of the marginal (i.e., the thermodynamic uncertainty relation) and is saturable. We illustrate the result with examples and we apply it to achieve the most accurate experimental estimation of dissipation associated to quantum annealing to date.
\end{abstract}

\maketitle

Entropy production is a central concept in non-equilibrium thermodynamics \cite{deGroot84Book}. It quantifies the degree of energy dissipation involved in a non-equilibrium process, and is also a measure of how far a system is driven away from equilibrium \cite{Schloegl66ZP191,Kawai07PRL98,Deffner10PRL105,Vaikuntanathan09EPL87,Campisi16JPA49a}. Furthermore when the system under study is a machine (natural or artificial) that performs some task, it is a prime quantity of interest for the understanding of the machine's functioning and for quantifying its efficiency. 
For this reason estimating entropy production is in the limelight of both biophysical research \cite{Roldan10PRL105,Seifert18PHYSA504,Li19NATCOMM10,Martinez19NATCOMM10,Horowitz20NATPHYS16,Skinner21PNAS118} and research on the thermodynamics of open quantum systems \cite{Deffner10PRL105,Parrondo09NJP09,Horowitz13NJP15,Timpanaro19PRL123,Landi20arXiv:2009.07668}. In particular,
quantifying entropy production can be  useful for understanding the mechanisms that lie at the basis of quantum information processing in quantum computers, and for learning how to master them, in an effort to improve their performance \cite{Buffoni20QST5,Solfanelli21arXiv:2106.04388}. 

Notably, however, in an experiment, classical or quantum, involving non-equilibrium processes in nano systems, one typically does not have direct experimental access to the entropy production $\sigma$ which is a global quantity pertaining to both the system and its environment. Instead one typically has access to some local experimentally accessible quantity pertaining to the system alone, which may be correlated with $\sigma$. A paradigmatic example is that of an open quantum system that is subject to some external driving, that performs work $W$ on it, while the system can exchange heat $Q$ with the surrounding environment at temperature $\beta_E$. Typically one does have only access to system observables, e.g., the  system energy variation $\Delta E= W-Q$, while having no direct access to the work $W$, the heat $Q$, nor the entropy production $\sigma=\beta_E Q + \beta \Delta E$ (here $\beta$ is the system initial energy).
Noisy intermediate-scale quantum computers belong to this category \cite{Preskill18QUANTUM2}, see also the experiments on NV centres reported in Ref. \cite{Hernandez21NJP23}. Notably, all the above quantities, i.e.,  $W,Q,\Delta E, \sigma$ are  stochastic quantities which can fluctuate from one realisation to the other of the same process \cite{Campisi11PRE83,Esposito09RMP81}. Thus the question naturally arises as to whether one can infer some information on the expectation of entropy production $\av{\sigma}$   based on sole information contained in the statistics of the measurable quantity $\Delta E$. 
Clearly, that is impossible if nothing is known about whether and how the measured quantity  (which from now on we will generally indicate with the symbol $\phi$), is correlated to $\sigma$. However, for a generic open system (classical or quantum) that starts in a state of thermal equilibrium, as in the case described above, it is known that the multivariate fluctuation theorem holds  \cite{Andrieux09NJP11,Campisi11PRE83}. In its simplest form, valid for driving protocols that are cyclic and time-reversal symmetric, it reads:
\begin{align}
\frac{p(\sigma,\phi)}{ p(-\sigma,-\phi)}=e^\sigma \, ,
\label{eq:biFT}
\end{align}
where $p(\sigma,\phi)$ is the probability distribution function for the  joint occurrence of $\sigma$ and $\phi$ in a single run of the process. The question is then, knowing that Eq. (\ref{eq:biFT}) holds, can one at least establish some bounds on $\av{\sigma}$, based on the sole knowledge of the marginal distribution $F(\phi)= \int d\sigma p(\sigma, \phi)$?

The answer is in the affirmative. A bound can be easily derived as follows. Take the logarithm of both sides of Eq. (\ref{eq:biFT}) and integrate over $p(\sigma,\phi)$, to obtain:
$
\langle \sigma \rangle =
D[p(\sigma,\phi)||p(-\sigma,-\phi)]
$
where $D[p||q]= \int d\sigma d \phi p (\ln p -\ln q)$ denotes the Kullback-Leibler divergence of $p$ with respect to $q$ \cite{ThomasBookInfoTheory}.
Since the latter contracts as information is discarded \cite{VanErven14TIT60}, it is:
\begin{align}
\av{\sigma}\geq D[F||\hat F] \,,
\label{eq:DAR}
\end{align}
where
\begin{align}
 D[F||\hat F] \doteq \int d\phi F(\phi) \ln \frac{F(\phi)}{F(-\phi)}
 \label{eq:D1[F]}
\end{align}
is the Kullback-Leibler divergence between $F(\phi)$ and its mirror image $\hat F(\phi)=F(-\phi)$.
Thus, if you have the full statistics of $\phi$, its degree of asymmetry (as quantified by $D[F||\hat F]$) 
provides a lower bound on the average dissipation. 

But Eq. (\ref{eq:DAR}) is not the only bound you can write down for $\av{\sigma}$. For example in Ref. \cite{Timpanaro19PRL123} it has ben shown that Eq. (\ref{eq:biFT}) implies the following relation:
\begin{align}
\langle \sigma \rangle 
 \geq h \left(\frac{\langle \phi \rangle}{\sqrt{\av{\phi^2}}}\right)
 ,\, h(x)\doteq 2 x\tanh^{-1}x.
\label{eq:TURtimpanaro2}
\end{align}
where $f(x)=\csch^2[g(x/2)]$, and $g(x)$ is the inverse of the function $x\tanh x $. 
Equation \ref{eq:TURtimpanaro2} is an instance of a so-called thermodynamic uncertainty relation (TUR), namely a relation expressing that higher precision comes with higher thermodynamic cost \footnote{A better nomenclature would perhaps be ``non-equilibrium uncertainty relations'' to not confuse them with the long-known  thermodynamic uncertainty relations expressing bounds on the product of fluctuations of thermodynamically conjugate equilibrium quantities (e.g., internal energy and temperature) \cite{Uffink99FP5}. The expression  thermodynamic uncertainty relation is also employed to denote the bound on the product of the fluctuations of mechanically conjugated quantities (i.e., position and velocity) of Brownian particles, in the context of the analogy between Brownian motion and Schr\"odinger's dynamics  \cite{Furth33ZP81,Peliti20arXiv2006.03740}. 
}.
 Such relations were first discovered in \cite{Barato15PRL114} and are currently under intense investigation, see Ref. \cite{Horowitz20NATPHYS16} for a perspective article on this topic.
In Ref. \cite{Buffoni19PRL122} we have in fact used Eq. (\ref{eq:TURtimpanaro2}) to experimentally estimate the amount of entropy production in a quantum computer, specifically a quantum annealer. That study allowed to corroborate the idea that  dissipation plays indeed an important role, not necessarily a detrimental one, in the process of quantum annealing. 

The remaining question is then whether it is possible to establish, on general grounds, whether one of the two bounds in Eqs. (\ref{eq:DAR},\ref{eq:TURtimpanaro2}) is tighter than the other. Below we shall prove that indeed the bound in Eq. (\ref{eq:DAR}) is tighter than the bound in Eq. (\ref{eq:TURtimpanaro2}):
\begin{align}
 D[F||\hat F]   \geq h \left(\frac{\langle \phi \rangle}{\sqrt{\av{\phi^2}}}\right)\, .
\label{eq:D>h}
\end{align}
We shall then illustrate this result with a mathematical example, a physical example (specifically the SWAP quantum heat engine  \cite{Quan07PRE76,Campisi15NJP17}), and finally apply it to to obtain an improved estimation of entropy production in a quantum annealing experiment. We shall also show that the bound $D[F||\hat F]$ can be saturated, and that happens when $\sigma$ and $\phi$ are delta-correlated. 

\emph{Proof of Eq. (\ref{eq:D>h})--} 
To prove Eq. (\ref{eq:D>h}) we ask ourselves which has smallest asymmetry $D[F||\hat F]$ among all probability density functions $F(\phi)$ that are normalised and have a given precision $P=\av{\phi}/\var{\phi}$. Introducing two Lagrange multipliers, $\lambda$ and $\gamma$, that amounts to find the minimum of the following functional:
\begin{align}
\mathcal D&[F,\lambda,\gamma] = \int d\phi F(\phi)\left[\ln \frac{F(\phi)}{F(-\phi)} -\frac{\gamma \phi}{\var{\phi}} -\lambda \right] \nonumber \\
&= \int d\phi F(\phi)\left[\ln F(\phi) - \ln\left( F(-\phi)e^{\frac{\gamma \phi}{\var{\phi}}+\lambda}\right) \right]\, . \nonumber 
\end{align}
Now, recalling that $\int dx f (\ln f-\ln g) \geq \int dx (g-f)$ and that the bound is saturated for $f\equiv g$  \cite{BalianBook1}, we see that $\mathcal D$ reaches its minimum (that is zero) for any $[\{F(\phi)\},\lambda,\gamma]$ such that:
\begin{align}
\frac{F(\phi)}{F(-\phi)}= \exp\left(\frac{\gamma \phi}{\var{\phi}}+\lambda\right)\, ,
\label{eq:FT-univariate_F}
\end{align}
Note that, for $\phi=0$, Eq. (\ref{eq:FT-univariate_F}) reduces to $e^\lambda=1$, implying that $\lambda=0$.
However, Eq. (\ref{eq:FT-univariate_F}) does not uniquely single out $F(\phi)$ and $\gamma$. Accordingly, each combination of $F(\phi),\gamma$ obeying Eq. (\ref{eq:FT-univariate_F}) (with $\lambda = 0$ and $\gamma$ chosen so that the constraint on precision be satisfied) corresponds to a local constrained extremum of $D[F||\hat F]$.
In order to find the absolute constrained minimum we proceed by introducing the rescaled Lagrange multiplier $\beta = \gamma/\var{\phi}$, and the rescaled normalised distribution
$
G(\phi)=F(\phi/\beta)/\beta
$
so that  $G$ obeys the univariate fluctuation relation 
$
G(\phi) =  G(-\phi) e^{\phi}.
$
Now we employ the following result: For a univariate probability distribution function $G$, obeying 
$
G(\phi) =  G(-\phi) e^{\phi}
$ it is:
\begin{align}
\av{\phi}_G \geq h\left( \av{\phi}_G/\sqrt{\av{\phi^2}_G} \right)\, .
\label{eq:TUR-univariate}
\end{align}
Here we have used the label $G$ to denote averages computed with respect to $G$. The proof of Eq. (\ref{eq:TUR-univariate}) is presented in the Supplementary Informations.
Using Eq. (\ref{eq:FT-univariate_F}) with $\lambda=0$, Eq. (\ref{eq:TUR-univariate}),  and noting that $\av{\phi^n}_G= \beta^n \av{\phi}$, we obtain
 \begin{align}
D&[F||\hat F]= \beta \av{\phi} = \av{\phi}_G \\ 
&\geq h\left( \av{\phi}_G/\sqrt{\av{\phi^2}_G} \right)= h\left( \av{\phi}/\sqrt{\av{\phi^2}} \right)\, . \nonumber
\end{align}
That means the minimum we are looking for is not smaller than $h( \av{\phi}/\sqrt{\av{\phi^2}})$, which concludes the proof.
It is worth adding that the bound is saturated by the distribution 
$
F_{\delta}(\phi)= p\delta (x-a) +(1-p)\delta (x+a)\, 
$
 ($\delta(\cdot)$ denotes Dirac delta function)
for which it holds $h(\av{\phi}/\sqrt{\av{\phi^2}})=(2p-1)\ln[p/(1-p)]=D[F_\delta||\hat{F}_\delta]$.

The fluctuation relation in the form of Eq. (\ref{eq:biFT}), applies when the underlying microscopic dynamics are time-reversal symmetric. That is not the case, generally, for quantum or classical systems subject to time-dependent forcing, for which the fluctuation relation takes on the form ${p(\sigma,\phi)}/{ \widetilde p(-\sigma,-\phi)}=e^\sigma$,
where $\widetilde p(\sigma,\phi)$, the so called backward pdf, denotes the joint pdf under the action of the time-reversed driving protocol \cite{Campisi11RMP83}. Equation (\ref{eq:DAR}) can be easily extended to include such cases, by replacing $\hat F$ with $\hat{\widetilde F}$, where ${\widetilde F}$ is the marginal of $\widetilde p$, and $\hat{\widetilde F}(\phi)={\widetilde F}(-\phi)$

\emph{The bound (\ref{eq:DAR}) is saturable.--}
The bound in Eq. (\ref{eq:DAR}) is saturated whenever $\sigma$ and $\phi$ are fully correlated, namely they are functionally dependent via some invertible and differentiable function $\phi=\varphi(\sigma)$: 
\begin{align}
p(\sigma,\phi)= \delta [\phi-\varphi(\sigma)] S(\sigma) 
\label{eq:fullCorr}
\end{align}
Note that by virtue of Eq. (\ref{eq:biFT}), the marginal distribution $S$ generally obeys the uni-variate fluctuation relation $S(\sigma)=S(-\sigma)e^\sigma$, and that in turn implies that $\varphi$ is antisymmetric: $\varphi(\sigma)=-\varphi(-\sigma)$. Integrating Eq. (\ref{eq:fullCorr}) in $d \sigma$ one gets $F(\phi)=S(\varphi^{-1}(\phi))/|\varphi'(\varphi^{-1}(\phi))|$, with $\varphi'$ denoting the derivative of $\varphi$. Noting that $\varphi'$ is an even function and using the change of variable $\phi=\varphi(\sigma)$, one finds
$
D[F||\hat F]= D[S||\hat S] = \av{\sigma} 
$. Clearly, in the case of full correlation, the statistics of $\phi$ contains full information on the statistics of $\sigma$. At variance with the TUR bound, Eq. (\ref{eq:TURtimpanaro2}), the asymmetry bound, Eq. (\ref{eq:DAR}) exploits those correlations fully by giving exactly the average value of dissipation.

\emph{Mathematical example.--} 
\begin{figure}[t]%
    \includegraphics[width=\linewidth]{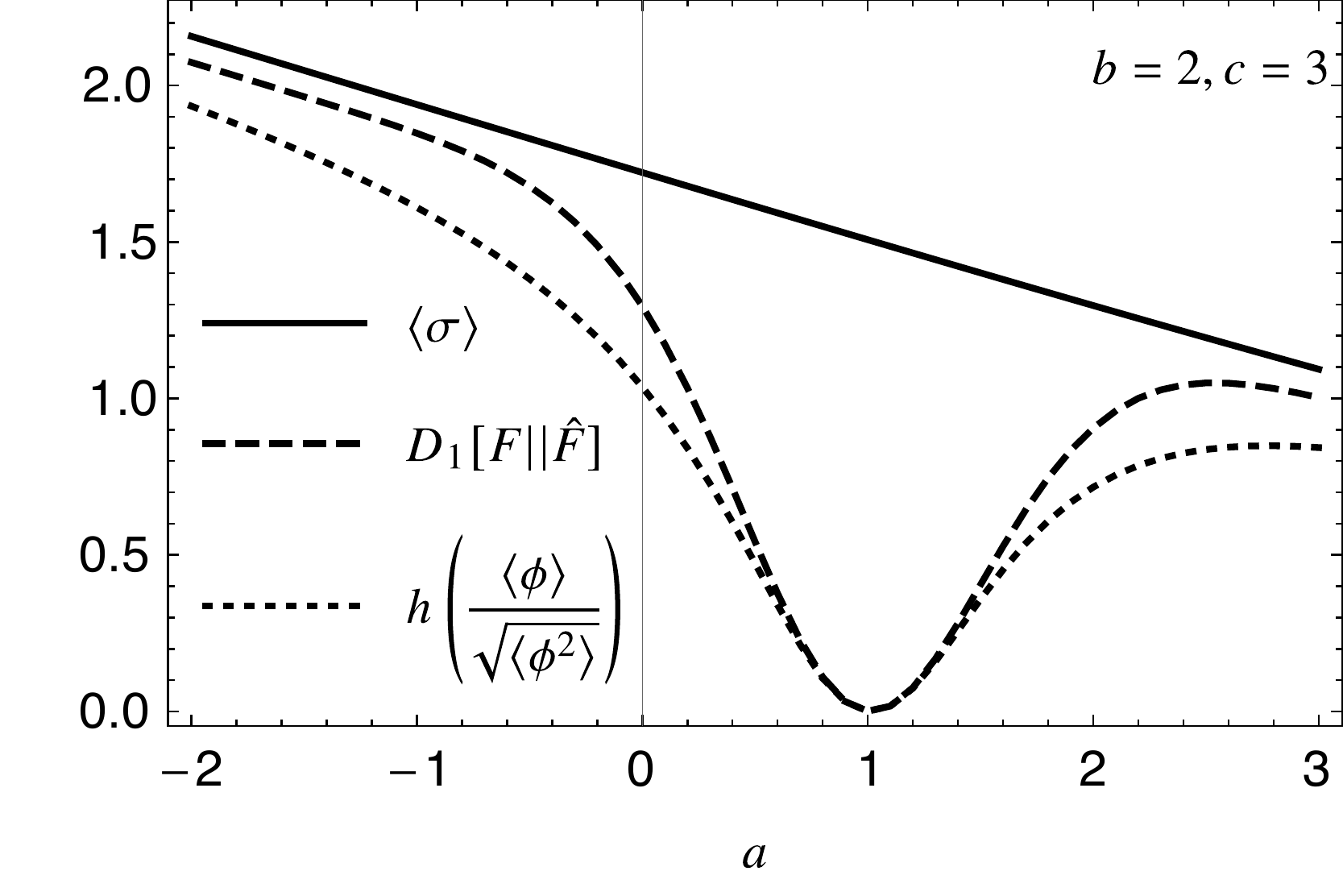} 
    \caption{Average dissipation $\av{\sigma}$, asymmetry bound $D[F||\hat F]$, Eq. (\ref{eq:DAR}), and TUR bound $h(\av{\phi}/\sqrt{\av{\phi^2}})$, Eq. (\ref{eq:TURtimpanaro2}), as a function of $a$, for the family in Eq. (\ref{eq:g-abc}), with fixed $b,c$.
    }
    \label{fig:Fig1}%
\end{figure}
Given some non-negative function $g(\sigma,\phi)$ consider the distribution
\begin{align}
p(\sigma,\phi)= n [g(\sigma,\phi)\theta(\sigma) + g(-\sigma,-\phi) e^\sigma\theta(-\sigma)] \, ,
\end{align}
which, by construction, obeys the fluctuation theorem, Eq. (\ref{eq:biFT}) (here $\theta$ denotes Heaviside step function and $n$ is the normalisation). Just for illustrative purposes  we consider the family
\begin{align}
g(\sigma,\phi)= \exp[-(\phi - a)^2 - c (\sigma - b)^2 - \phi \sigma]\, ,
\label{eq:g-abc}
\end{align}
parametrised by the real numbers  $a,b,c$.
Fig. \ref{fig:Fig1} shows $\av{\sigma}$, the TUR bound $h(\av{\phi}/\sqrt{\av{\phi^2}}$),
and the asymmetry bound $D[F||\hat{F}]$ as a function of $a$ for fixed $b,c$ (qualitatively similar plots would be obtained for different values of $b,c$). The figure shows that the bound in Eq. (\ref{eq:DAR}) is tighter than the TUR bound in Eq. (\ref{eq:TURtimpanaro2}) as expected. 
We observe the salient fact that $D[F||\hat{F}]$ not only is a good estimator of $\av{\sigma}$ in the vicinity of equilibrium (i.e., for small $\av{\sigma}$), but may be good as well far away from equilibrium (i.e., for large $\av{\sigma}$). Good performance for low dissipation  $\av{\sigma}$ can be easily understood on the basis of $\av{\sigma} \geq D[F||\hat{F}] \geq 0$, which implies that $D[F||\hat{F}]$ goes to zero as  $\av{\sigma}$ decreases, and so does their difference. 
However good performance is achieved as well whenever $F$ is very asymmetric (which can only occur at high dissipation). This so because  asymmetry of $F$ is a quantifier of correlations \footnote{Note that in absence of correlations, i.e. when the joint probability reads $P(\sigma,\phi)=F(\phi)S(\sigma)$, Eq. (\ref{eq:biFT}) implies $F(\phi)=F(-\phi)$, hence vanishing asymmetry $D[F||\hat{F}]$}, and correlations is the ingredient that allows to lift the bound on average dissipation above its lowest value, i.e., 0.
The dips in the plots correspond to cases where the correlations drop down and the $\phi$ distribution is highly (though not perfectly) symmetric: When one records a symmetric $F$, without further information, one cannot rule out independence of $\phi$ and $\sigma$, and this is the reason why the bounds cannot be lifted from their lowest value, $0$, in that case.

\emph{Physical example.--} As a physical example we consider  a prototypical quantum heat engine, namely the two-qubit, two-stroke SWAP engine \cite{Quan07PRE76,Campisi15NJP17,Solfanelli21arXiv:2106.04388}. Two qubits with Hamiltonians  $H_i= \hbar \omega_i\sigma_i^z /2$, $i=1,2$, are prepared each in a thermal state at inverse temperature, $\beta_1$ and $\beta_2$, respectively (here $\omega_i$ is qubit $i$ resonant frequency and $\sigma_i^z$ is qubit $i$ $z$-Pauli matrix).
In the first stroke the SWAP unitary $U_{\text{SWAP}}$ is applied to the two qubits. In the second stroke the two qubits are allowed each to relax back to their initial thermal state by contact with heat reservoirs at inverse temperatures $\beta_1$ and $\beta_2$. Projective measurements of each qubit energy are performed right before and after the SWAP is applied. Calling $\Delta E_i$ the energy change of qubit $i$ in an individual run of the process, the joint pdf $q(\Delta E_1,\Delta E_2)$ of occurrence of $\Delta E_1$ and $\Delta E_2$, reads \cite{Campisi15NJP17}
\begin{align}
&q(\Delta E_1,\Delta E_2)=  \frac{\delta(\Delta E_1+\Delta E_2)}{Z_1Z_2} \times
\\ 
&\left[ 2\delta(\Delta E_2) \cosh\left(\frac{\sigma}{2}\right) 
+ \delta(\Delta E_2-\omega_2) e^{-A}
+ \delta(\Delta E_2+\omega_2) e^{A} \right], \nonumber
\end{align}
where $Z_i=\Tr e^{-\beta_i H_i}=2\cosh[ \hbar \omega_i/2 ]$ is the partition function of qubit $i$ and $A=\hbar ( \beta_1 \omega_1 -   \beta_2 \omega_2)$.
Note that $\Delta E_1$ and $\Delta E_2$ are functionally dependent $\Delta E_2=-\Delta E_1$, and so $\sigma= \beta_1 \Delta E_1 + \beta_2 \Delta E_2$, is functionally dependent on $\Delta E_1\doteq \phi$, i.e., $\phi = \sigma/(\beta_1 -\beta_2)$.
Thus, the joint probability $p(\sigma,\phi)$ obeys the fluctuation theorem  (\ref{eq:biFT}) and is of the form (\ref{eq:fullCorr}), therefore the bound ($\ref{eq:DAR}$) is always saturated in this case, regardless of how far the engine operates away from equilibrium, that is regardless of how much its efficiency, $\eta=1-\omega_2/\omega_1$, deviates from the Carnot efficiency, $\eta_C = 1- \beta_1/\beta_2$. This should be contrasted to the TUR bound (\ref{eq:TURtimpanaro2}) which has been shown to saturate only at the Carnot point where the engine is in stall \cite{Campisi15NJP17}.

\emph{Quantum annealing experiment.-- }
To demonstrate the practical relevance of our results we apply it to address the problem of estimation of dissipation in quantum computing, specifically in quantum annealing. Without entering the details of what a quantum annealer is, how it works and how it is used in practice to solve optimisation problems \cite{Das08RMP80}, it suffices here to mention that a quantum annealer is a lattice of programmable superconducting qubits, with tunable interactions and local fields, which can be prepared and measured in a given eigenbasis. It thus implements a driven quantum spin network on a low temperature microchip.

The main problem in estimating dissipation associated to quantum annealing  is that in order to obtain it experimentally, one would need to perform measurements of the energy that flows into its surrounding environment, which with current technology is practically impossible.
Due to the extreme complexity of the system, performing a faithfull ab-initio numerical simulation of its quantum open dynamics, is also extremely challenging, and currently is an open problem, which makes as well any numerical attempt not a viable option.  The only practical way to estimate entropy production is to use the available partial experimental information, to get at least a lower bound. In Ref.  \cite{Buffoni20QST5} we have done so using the TUR  bound (\ref{eq:TURtimpanaro2}), here we perform a new experiment, using the asymmetry bound (\ref{eq:DAR}), and demonstrate an improvement in the estimate.

Using  D-Wave systems Leap Service \footnote{\url{https://cloud.dwavesys.com/leap/}.}, we remotely run an experiment on the D-Wave 2000Q lower-noise quantum processor. We set up the experimental parameters so that the dynamics of the processor are well described by  the time dependent spin-chain Hamiltonian
\begin{equation}
    H(t) = [1-s(t)]\sum _i^L \sigma ^x_i + s(t)\left( \sum _i^L  \sigma ^z_i + \sum _{i}^{L-1} \sigma ^z_i \sigma ^z_{i+1}\right)
    \label{eq:H}
\end{equation}
with $s(t)$ the so called annealing parameter, and $L$ the chain length. In our experiment $s(t)$ ramps down linearly  from $s=1$ to $s=\bar s$, in the time span $(0,\tau/2)$, and ramps up back linearly to $s=1$ in  the time span $(\tau/2,\tau)$, thus realising a so called reverse annealing schedule \cite{Buffoni20QST5}. For fixed reverse annealing parameters $(\bar s, \tau)$, we repeat the annealing schedule $N$ times, with the  processor being prepared each time in an eigenstate of the initial Hamiltonian $H(0)$ corresponding to some energy $E_n$, with the according Gibbs probability $p_n= e^{-\beta E_n}/Z$. At time $\tau$, we record the final energy of the processor $E_m$, and so construct the statistics $F(\Delta E)$ of the energy change $\Delta E = E_m-E_n \doteq \phi$ of the processor. Our working assumption is that such statistics is the marginal of a joint distribution $p(\sigma,\phi)$ obeying the fluctuation relation, as discussed in Ref. \cite{Buffoni20QST5}, where $\sigma = \beta \Delta E + \beta_E Q$ with $\beta_E$ the inverse temperature of the processor environment and $Q$ the heat it receives from the environment \footnote{The code we employed to run our experiments using D-Wave's hardware is publicly available at \url{https://github.com/Buffoni/dwave_notebooks/}.}.

\begin{figure}[t]%
\includegraphics[width=\linewidth]{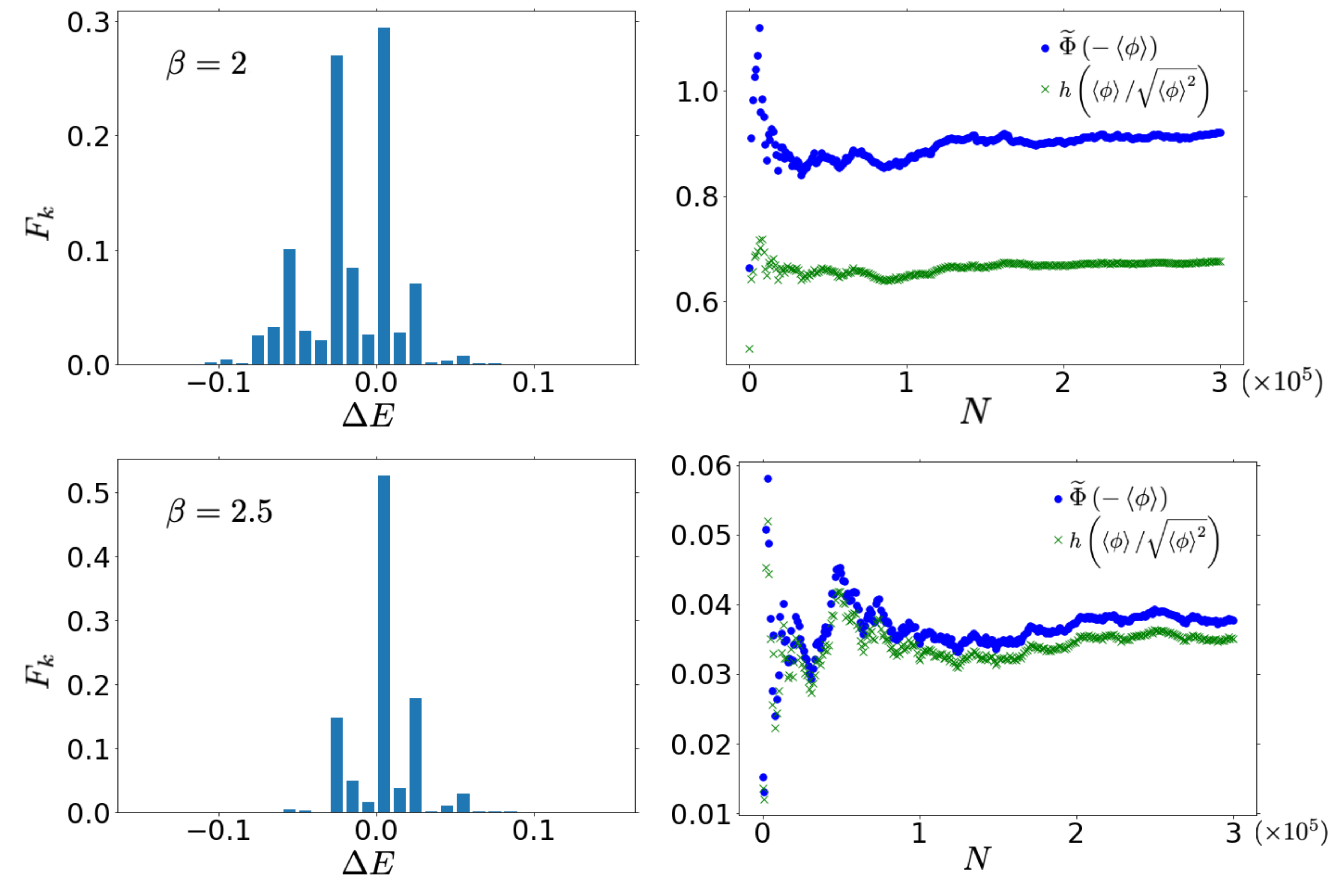}
    \caption{Left panels: Experimental discrete probabilities $F_k$ for $\beta=2$ (top) and $\beta=2.5$ (bottom), the bin width is $d=0.01$ and $N= 3 \times 10^5$. Right panels: Kullback bound $\widetilde \Phi(-\av{\phi})$, Eq (\ref{eq:Kineq}) and TUR bound, Eq. (\ref{eq:TURtimpanaro2}), as functions of sample size $N$, for $\beta=2$ (top) and $\beta=2.5$ (bottom). For all plots it is $\bar s =0.3 , \tau =4 \mu s , L = 300$.}
    \label{fig:Fig2}%
\end{figure}

Figure \ref{fig:Fig2}, left panels, shows the discrete probabilities $F_k\doteq \text{Prob}[(k-1/2) d < \Delta E \leq (k+1/2)d]$ to find the energy change $\Delta E$ in the $k$'th bin of width $d$, obtained for two distinct values of $\beta$ and fixed $\bar s, \tau, N, L$.
Note that our $F$ displays a non-symmetric support, implying that its asymmetry $D[F||\hat F]$ is not defined.
This does not mean that Eq. (\ref{eq:DAR}) is useless in practical situations. On the contrary, we note that, by virtue of the Kullback inequality \cite{Kullback59Book}, it is:
\begin{align}
\av{\sigma} \geq D[F||\hat F] \geq \sup_v [ - v \av{\phi} -\Phi(v)] = \widetilde{\Phi}(-\av{\phi})
\label{eq:Kineq}
\end{align}
where $\Phi(v)=\ln \av{e^{v\phi}}$ is the cumulant generating function associated with $F$, and $\widetilde{\Phi}(-\av{\phi})$ is its Legendre-Fenchel transform evaluated at $-\av{\phi}$. The latter is well defined in the case of non-symmetric support (as long as the latter is not fully contained in one semi-axis).
Figure \ref{fig:Fig2}, right panels, shows how our estimates of $h(\av{\phi}/\sqrt{\av{\phi^2}})$ and 
$\widetilde{\Phi}(-\av{\phi})$ behaved as the size $N$ of our sample increased. Note that the bound  $\widetilde{\Phi}(-\av{\phi})$ is larger than the TUR bound $h(\av{\phi}/\sqrt{\av{\phi^2}})$ for all $N$'s. Also note that their difference is larger for the more asymmetric distribution, and that the estimate of $\widetilde{\Phi}(-\av{\phi})$ appears to be subject to larger absolute fluctuations than $h(\av{\phi}/\sqrt{\av{\phi^2}})$, while relative fluctuations appear unaltered. The present method is accordingly more powerful and equally robust as compared to the TUR bound \ref{eq:TURtimpanaro2}. Indeed it is safe to say that our method for estimating entropy production in a quantum annealer is the most accurate put forward so far, and is as well of simple and immediate applicability.

\emph{Conclusions.--} 
According to thermodynamic uncertainty relations, the average dissipation incurred in a non-equilibrium thermodynamic process  is lower bounded by an increasing function of precision of any stochastic variable that enters the fluctuation relation. The average dissipation is also lower bounded by the degree of asymmetry of the distribution of any such variable. Here we proved that the latter is a tighter bound than the former and it is saturable, hence it can be used to improve estimates of average dissipation in nano systems and devices. The result was formally proved, illustrated with a generalisation, a mathematical and a physical example, and finally applied to obtain the best to date experimental estimate of dissipation during operation of a quantum annealer.

\appendix
\section{Supplementary material}
Here we prove that, if $f(x)$ is a univariate distribution over  $\mathbb{R}$ obeying 
\begin{align}
f(x)=e^x f(-x)\, ,
\label{eq:FTfx}
\end{align}
then 
\begin{align}
\av{x} \geq h(\av{x}/\sqrt{\av{x^2}})
\label{eq:TURuniv_f(x)}
\end{align}
The proof follows closely the proof of Eq. (\ref{eq:TURtimpanaro2}) for bivariate distribution obeying Eq. (\ref{eq:biFT}) presented in Refs.  \cite{Zhang19arXiv,VanVu20JPA53}.
First let us introduce the probability distribution function $q(x)=(1+e^{-x})f(x)$ defined on the positive $x$-axis, $\mathbb{R}^+$.  From Eq. (\ref{eq:FTfx}) it follows
\begin{align}
\av{x^{2n}} = \av{x^{2n}}_q\, , \quad  \av{x^{2n+1}} = \av{x^{2n+1}\tanh(x/2)}_q
\end{align}
where $\av{\cdot}$ denote the average over $f(x)$, $\av{\cdot}_q$ denote average over $q(x)$, and $n$ is a non-negative integer. 
Using the Cauchy-Schwartz inequality we have:
\begin{align}
\av{x}^2 = \av{x \tanh(x/2)}_q^2 \leq 
\av{x^2} \av{\tanh^2(x/2)}_q %= \av{x^2} \av{\tanh^2(x/2)}
\end{align}
Now define $r(x)= \tanh^2(x/2)$, $k(x)\doteq x \tanh(x/2)$, $g\doteq k^{-1}$ and $w \doteq r \circ g$. The function $w$ is strictly concave $w''>0$, therefore, using Jensen's inequality
\begin{align}
&\av{\tanh^2(x/2)}_q = \av{r(x)}_q = \av{r(g(k(x)))}_q  = \av{w(k(x))}_q \nonumber\\
& \leq  w(\av{k(x)}_q) = r(g(\av{k(x)}_q)) = r(g(\av{x})) 
\end{align}
where in the last equality we used $\av{k(x)}_q= \av{x \tanh(x/2)}_q=\av{x}$. Combining the above two equations, and using the definition of $r$ we get:
\begin{align}
\av{x}^2 \leq \av{x^2}\,  r(g(\av{x}))  = \av{x^2} \tanh^2\left(\frac{g(\av{x})}{2}\right)
\end{align}
Dividing both sides by $ \av{x^2}$, taking their square root, applying then the increasing function $\tanh^{-1}$, multiplying them by 2 and finally applying $k=g^{-1}$, one obtains Eq. (\ref{eq:TURuniv_f(x)}).

\end{document}